\documentclass[english,floatfix,twocolumn,superscriptaddress,prb,aps]{revtex4}
\usepackage[T1]{fontenc}
\usepackage[latin1]{inputenc}
\usepackage{graphicx}
\usepackage{amssymb}
\usepackage{amsmath}
\usepackage{bm}
\usepackage{babel}
\usepackage{color}
\makeatother
\renewcommand{\vec}[1]{\mathrm{\mathbf{#1}}}
\begin{document}

\title{Graphene Mode-Locked Ultrafast Laser}
\author{Z. Sun$^1$, T. Hasan$^1$, F. Torrisi$^1$, D. Popa$^1$, G. Privitera$^1$, F. Wang$^1$,F.Bonaccorso$^1$, D. M. Basko$^2$, A. C. Ferrari}\thanks{acf26@eng.cam.ac.uk}
\affiliation{Department of Engineering, University of Cambridge, Cambridge, UK\\
$^2$LPMMC,Universit\'e Joseph Fourier and CNRS, Grenoble, France}
\begin{abstract}
Graphene is at the center of a significant research effort. Near-ballistic transport at room temperature and high mobility make it a potential material for nanoelectronics. Its electronic and mechanical properties are also ideal for micro and nanomechanical systems, thin-film transistors and transparent and conductive composites and electrodes. Here we exploit the optoelectronic properties of graphene to realize an ultrafast laser. A graphene-polymer composite is fabricated using wet-chemistry techniques. Pauli blocking following intense illumination results in saturable absorption, independent of wavelength. This is used to passively mode-lock an Erbium-doped fibre laser working at 1559nm, with a 5.24~nm spectral bandwidth and $\sim$460~fs pulse duration, paving the way to graphene-based photonics.
\end{abstract}
\maketitle

Ultrafast laser sources have many potential applications, ranging from basic research and metrology to telecommunications, medicine and materials processing\cite{dausingertiap04,kellernat03,schutzecellmol98}. The majority of ultrashort lasers employ a mode-locking technique, whereby a non-linear optical element - called saturable absorber - turns the laser continuous wave output into a train of ultrashort optical pulses. Semiconductor Saturable Absorber Mirrors (SESAMs) currently dominate passive mode-locking\cite{kellernat03,xiangieee02,steinmeyersci99}. However, these have a narrow tuning range (tens of nm), and require complex fabrication and packaging\cite{kellernat03}. A simpler and cost-effective alternative relies on Single Wall Carbon Nanotubes (SWNTs) \cite{rozhinapl06,scardaciadvmater08,wangnatnano08}, where the working wavelength is defined by choosing the SWNT diameter (\textit{i.e.} bandgap)\cite{rozhinapl06,scardaciadvmater08}. Tunability is possible by combining SWNTs with a wide diameter distribution\cite{wangnatnano08}. However, when operating at a particular wavelength, the SWNTs not in resonance are not used, and contribute unwanted insertion losses, compromising device-performance. Novel nonlinear materials with broadband absorption are therefore required for wideband, tunable operation.

The linear dispersion of the Dirac electrons in graphene offers the ideal solution\cite{geimnatmater07}: for any excitation energy there will always be an electron-hole pair in resonance. Due to the ultrafast carrier dynamics\cite{breusingprl09,sunprl08,seibertprb90} and large absorption of incident light per layer ($\alpha_1=2.3\%$~ \cite{nairsci08,CasiraghiNL}), graphene should behave as a fast saturable absorber over a wide spectral range. Compared to SESAMs and SWNTs, graphene saturable absorbers would not need bandgap engineering or chirality/diameter control to optimize device performance. Here, we demonstrate an ultrafast fibre laser mode-locked at $\sim1.5\mu$m, the most common optical telecommunications wavelength, using single layer graphene (SLG) and few layer graphene (FLG) flakes.

For ease of integration and stability, we incorporate the flakes into a host polymer matrix. For the composite preparation, we process graphite in water without functionalization\cite{hernandez}. This allows us to retain the electronic structure of pristine graphene in the resulting exfoliated SLG and FLG flakes\cite{hernandez}. We employ bile salts in order to obtain a stable, higher concentration of SLG and FLG than in previous non-aqueous dispersions\cite{hernandez}. These amphiphilic molecules, with a hydrophobic and a hydrophilic side\cite{mukerjeejpsci74}, disperse graphene in water by physical adsorption on its surface. In contrast to linear chain surfactants, e.g. sodium dodecylbenzene sulfonate (SDBS) widely used for SWNTs, the flat molecular structure of sodium deoxycholate (SDC), used here, efficiently disperses SLG, FLG and graphitic flakes. Polyvinyl alcohol (PVA) is chosen as host polymer for its mechanical properties and solvent compatibility. The dispersions are mixed with PVA via ultrasonication. The solvent is then evaporated to obtain free-standing composites $\sim$50$\mu$m thick. This simple, wet-chemistry approach is scalable and, more importantly, allows easy integration into a range of photonic systems (see Methods).

\begin{figure*}
 \centerline{\includegraphics[width=175mm]{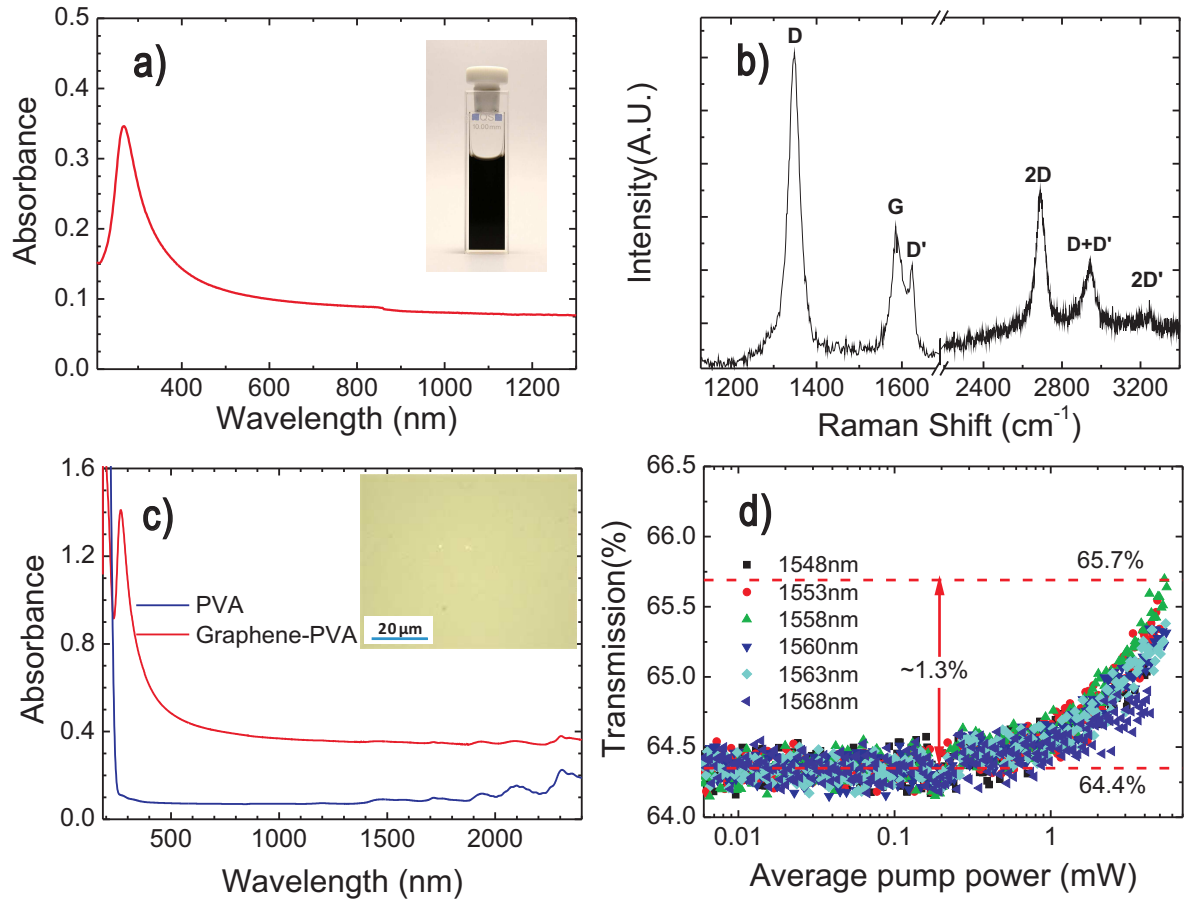}}
 \caption{\label{fig1}(a)Absorption spectrum of a stable, 10\% diluted, graphene dispersion. The insert shows a photograph of the undiluted dispersion.(b)Raman spectrum measured at 514.5nm for a representative flake.(c)Absorption spectra of graphene-PVA composite and a reference PVA film. Insert: micrograph of a graphene-PVA composite.(d)Typical transmission of the composite as a function of average pump power for six excitation wavelengths.}
\end{figure*}

We first characterize the dispersions obtained from ultrasonication, followed by centrifugation to remove large residual graphitic particles. Since we need to integrate our composites in transmissive mode in a fibre laser cavity, flakes with dimensions $>1\mu$m are removed to avoid optical scattering losses\cite{bohrenbook}. Fig.\ref{fig1}(a) shows a photograph of a dispersion after centrifugation and its absorption spectrum. This is mostly featureless, as expected\cite{nairsci08,CasiraghiNL,abergelprb07,hernandez,heinz}. The peak in the UV region is a signature of the $\pi$ plasmon\cite{eberlein}. Using the experimentally derived absorption coefficient of 1390 L.g$^{-1}$.m$^{-1}$ at 660nm (1.88 eV)\cite{hernandez}, we estimate $\sim$0.08g/L of graphitic material in the centrifuged dispersion.

The decanted dispersion mostly contains sub-micrometer flakes. A typical Raman spectrum of such flakes measured at 514.5nm is plotted in Fig.\ref{fig1}(b). Besides the G and 2D peaks, this has significant D and D' intensities, and the combination mode D+D'$\sim$2950 cm$^{-1}$. The G peak corresponds to the E$_{2g}$ phonon at the Brillouin zone centre. The D peak is due to the breathing modes of sp$^2$ rings and requires a defect for its activation by double resonance (DR)\cite{ACFRaman,Ferrari00,tk}. The 2D peak is the second order of the D peak. This is a single band in monolayer graphene, whereas it splits in four in bi-layer graphene, reflecting the evolution of the band structure\cite{ACFRaman}. The 2D peak is always seen, even when no D peak is present, since no defects are required for the activation of two phonons with the same momentum, one backscattering from the other. DR can also happen as intra-valley process, i.e. connecting two points belonging to the same cone around \textbf{K} or \textbf{K'}. This gives rise to the D' peak. The 2D' is the second order of the D' peak. The very large intensity of the D peak in Fig.1b is not due to the presence of a large amount of structural defects, otherwise it would be much broader, and G, D' would merge in a single band\cite{Ferrari00}. We rather assign it to the edges of the sub-micrometer flakes we produce. We note that the 2D band, although broader than in pristine graphene\cite{ACFRaman}, is still fitted by a single lorentzian lineshape. This implies that, even if the flakes are multi-layers, they are electronically almost decoupled and behave, to a first approximation, like a collection of single layers, retaining the Dirac fermions linear dispersion\cite{latil}.

The micrograph of the composite shown in the insert of Fig.\ref{fig1}(c) confirms the homogeneous distribution of the flakes. This is vital to reduce Mie scattering, that could be caused by flakes of dimensions comparable to $1.5\mu m$, the device operation wavelength\cite{bohrenbook}. The absorption spectra of the graphene-PVA composite and a reference PVA film are presented in Fig.\ref{fig1}(c). The absorption of the flakes is featureless with the characteristic UV plasmon peak, while the host polymer only contributes a small background for longer wavelengths.

Fig.\ref{fig1}(d) plots the measured  transmission as a function of average pump power at six different wavelengths (using a probe laser with pulse width 580~fs, as detailed in Methods). At a relatively low input power level, the transmission, $\tau$, is almost independent of pump power. However, $\tau$ increases by $\Delta\tau=1.3\%$ due to absorption saturation when the incident average power is raised to 5.35~mW (corresponding to a peak power density of 266~MW/cm$^2$) at 1558~nm. Further increase in transmission is feasible, but limited to our maximum available pump intensity. The saturable absorption of the composite is clear for all pump wavelengths. This indicates that our composite can be used over a broad spectral range, unlike SESAMs\cite{kellernat03}. The transmission change is similar to SESAMs\cite{kellernat03}. The non-saturable insertion loss ($34.3\%$) is larger than that of SESAMs\cite{kellernat03}, but comparable to SWNTs\cite{wangnatnano08}. Note that, for fibre lasers with relatively large single round-trip gain coefficient, such non-saturable losses are tolerable\cite{wangnatnano08}. Further decrease in non-saturable insertion loss is expected when the device is completely saturated. To estimate the non-saturable loss due to coupling between the two fibre connectors, a reference $\sim50\mu m$ PVA composite is used in place of the graphene-PVA composite. The transmission of the packaged pure PVA is $\tau_0=82.6\%$. Given the pulse repetition rate of 38.83~MHz, we can estimate the density of photons absorbed per pulse at our maximum pump power to be $2.3\times 10^{14}\:\mbox{cm}^{-2}$. The average number of layers $N$ participating in the absorption is estimated from $\tau=\tau_0(1-\alpha_1)^N$, where $\alpha_1=2.3\%$ is the absorption per layer, yielding $N\sim11$ (see Methods). Thus, the photon density absorbed per pulse per layer is $2.2\times 10^{13}\:\mbox{cm}^{-2}$, and $\Delta\tau=1.3\%$ translates into a relative change in the absorption per layer $\Delta\alpha_1/\alpha_1=-8.2\%$.
\begin{figure}
 \centerline{\includegraphics[width=90mm]{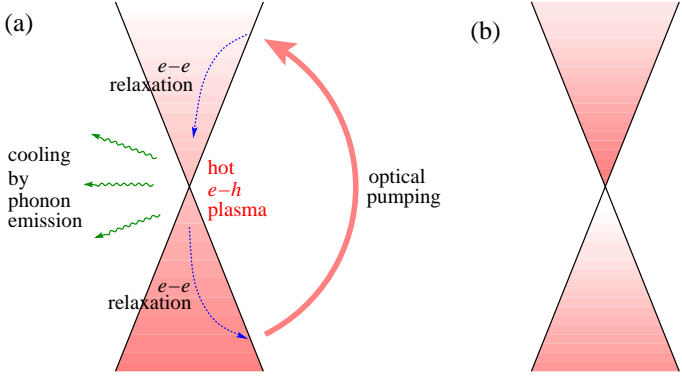}}
 \caption{\label{fig2} Schematic illustration of photoexcited electron kinetics in graphen. The intensity of the color shading represents the electron population, in the case of (a)~efficient interband relaxation and (b)~inefficient interband relaxation.}
\end{figure}

To understand the origin of the observed saturation and give its theoretical estimate, let us discuss the photoexcited carrier kinetics in graphene (Fig.~\ref{fig2}). Optical interband excitation in SLG and FLG by an ultrashort optical pulse produces a non-equilibrium carrier population in the valence and conduction bands. In time-resolved experiments~\cite{breusingprl09,kampfrathprl05} two relaxation time scales are typically seen. A fast one, $\sim{100}\:\mbox{fs}$, usually associated with carrier-carrier intraband collisions and phonon emission, and a slow one, on a picosecond scale, corresponding to electron interband relaxation and cooling of hot phonons~\cite{kampfrathprl05,lazzeriPRL}. Several points should be noted here. (i)~In our experiment, SLG and FLG flakes are incorporated in a polymer matrix, so we assume graphene phonons to efficiently give away their energy to the matrix and neglect hot phonon effects. (ii)~The rate of carrier-carrier collisions grows with carrier density, i.~e., pump power. We assume the collisions to be very fast.(iii)~Carrier-carrier collisions preserve the total electronic energy and thus cannot change the effective electronic temperature.

A quantitative treatment of the electron-hole dynamics would require the solution of the kinetic equation for electron and hole distribution functions $f_e(\vec{p})$ and $f_h(\vec{p})$, $\vec{p}$~being the momentum counted from the Dirac point. Here we present an estimate. Let us assume that relaxation times are shorter than the pulse duration, so during the pulse the electrons reach a stationary state, that collisions put electrons and holes in thermal equilibrium at an effective temperature $T_{eff}$, and take undoped samples:
\begin{equation}\label{FermiDirac=}
f_e(\vec{p})=f_h(\vec{p})=\frac{1}{e^{vp/T_{eff}}+1}\,
\end{equation}
where $v$~is the velocity of the Dirac electrons in SLG. The populations determine the electron and hole densities $n_{e,h}$ and the total energy density $\mathcal{E}$ (counted from the energy of the un-doped sample at zero temperature):
\begin{equation}\begin{split}
&n_{e,h}=4\int\frac{d^2\vec{p}}{(2\pi)^2}\,f_{e,h}(\vec{p}),\\
&\mathcal{E}=4\int\frac{d^2\vec{p}}{(2\pi)^2}\,vp
\left[f_e(\vec{p})+f_h(\vec{p})\right],\label{Edef=}
\end{split}\end{equation}
the factor 4 accounts for valley and spin degeneracy. The populations also determine the reduction of photon absorption per graphene layer for a given laser energy $E_L$, due to Pauli blocking, by a factor $1+\Delta\alpha_1/\alpha_1=[1-f_e(\vec{p})][1-f_h(\vec{p})]$. In our measurements, $vp=E_L/2=0.4\:\mbox{eV}$ and $1+\Delta\alpha_1/\alpha_1\sim0.92$, implying $T_{eff}\sim{0}.13\:\mbox{eV}$.

To obtain a theoretical estimate for $T_{eff}$, let us assume that during the pulse electrons and holes are injected at the energy $\epsilon_{in}=E_L/2=0.4\:\mbox{eV}$ at a constant rate,
\begin{equation}\begin{split}
&\left.\frac{dn_e}{dt}\right|_{pump}=
\left.\frac{dn_h}{dt}\right|_{pump}=J_{in}\\ &=
\frac{2.2\cdot{10}^{13}\:\mbox{cm}^{-2}}{0.58\:\mbox{ps}}
= 3.8\cdot{10}^{13}\:\mbox{cm}^{-2}\mbox{ps}^{-1}.\label{Jin=}
\end{split}\end{equation}
This corresponds to pumping the energy at a rate:
\begin{eqnarray}
&&\left.\frac{d\mathcal{E}}{dt}\right|_{pump}=2J_{in}\epsilon_{in}
= 3.0\cdot{10}^{13}\:\mbox{cm}^{-2}\mbox{ps}^{-1}\mbox{eV}.
\end{eqnarray}
The electronic energy density~$\mathcal{E}$ can be decreased by phonon emission. Neglecting hot phonon effects, for the Fermi-Dirac distribution~(\ref{FermiDirac=}) the cooling rate can be calculated as (see Methods):
\begin{eqnarray}
&&\left.\frac{d\mathcal{E}}{dt}\right|_{ph}=
-\frac{\lambda_\Gamma\omega_\Gamma^4}{4\pi{v}^2}\,\mathcal{I}(2T_{eff}/\omega_\Gamma)
-\frac{\lambda_K\omega_K^4}{4\pi{v}^2}\,\mathcal{I}(2T_{eff}/\omega_K)\nonumber\\
&&\qquad\qquad{}-\frac{7\pi^3}{30}\,L_\mathrm{ac}v^2\left(\frac{T_{eff}}{v}\right)^5,
\label{cooling=}\\
&&\mathcal{I}(y)=y^3\int\limits_0^\infty\frac{|x^2-1/y^2|\,dx}{e^{2/y}+2e^{1/y}\cosh{x}+1}.
\end{eqnarray}
The first two terms in Eq.~(\ref{cooling=}) correspond to emission of optical phonons with wave vectors near $\bf{\Gamma}$ and \textbf{K}, respectively. Their frequencies $\omega_\Gamma,\omega_K$ are assumed to be independent of wave vector (Einstein model). The dimensionless coupling constants $\lambda_\Gamma$ and $\lambda_K$ are defined in Ref.~\onlinecite{Basko2008}. The third term in Eq.~(\ref{cooling=}) corresponds to emission of longitudinal acoustic phonons, with frequency taken proportional to their momentum (Debye model). The length $L_\mathrm{ac}$ in Eq.~(\ref{cooling=}) is expressed in terms of deformation potential, $D_0$, carbon atom mass, $M$, and unit cell area, $A_\mathrm{u.c.}$, as $L_\mathrm{ac}=D_0^2A_\mathrm{u.c.}/(2Mv^3)$. For $D_0=10\:\mbox{eV}$, $A_\mathrm{u.c.}=5.24\:\mbox{\AA}^2$, $M=2.00\times{10}^{-23}\:\mbox{g}= {2}.88\times{10}^3\:(\mbox{eV}\cdot\mbox{\AA}^2)^{-1}$, $v=10^8\:\mbox{cm}/\mbox{s}={7}\:\mbox{eV}\cdot\mbox{\AA}$, we obtain $L_\mathrm{ac}=0.26\times{10}^{-3}\:\mbox{\AA}$. We take $\omega_\Gamma=0.20\:\mbox{eV}$, $\lambda_\Gamma=0.03$, $\omega_K=0.15\:\mbox{eV}$, $\lambda_K=0.1$~\cite{baskoCM}. These give the time of optical phonon emission by the photoexcited electron to be $[(\lambda_\Gamma/2)(E_L/2-\omega_\Gamma)+(\lambda_K/2)(E_L/2-\omega_K)]^{-1}={40}\:\mbox{fs}$~\cite{Basko2008}.

From the thermal balance equation,
\begin{equation}
\left.\frac{d\mathcal{E}}{dt}\right|_{pump}+\left.\frac{d\mathcal{E}}{dt}\right|_{ph}=0,
\end{equation}
we obtain a theoretical estimate $T_{eff}=0.20\:\mbox{eV}$, the dominant contribution to cooling coming from optical phonon emission. Given this value of $T_{eff}$, and using the Fermi-Dirac distribution~(\ref{FermiDirac=}), we obtain that the absorption of photons with energy $E_L=0.8\:\mbox{eV}$ should be reduced by up to $1+\Delta\alpha_1/\alpha_1=(1-f_e)(1-f_h)=0.78$. This corresponds to $\Delta\alpha_1/\alpha_1=-22\%$, while the measured one is $\Delta\alpha_1/\alpha_1=-8.2\%$. Thus, having assumed efficient carrier-carrier relaxation (both intraband and interband) and efficient phonon cooling, with the main bottleneck being the transfer of energy from the electrons to the graphene phonons, we obtain a higher saturation level than the measured one. On one hand, this points to other mechanisms contributing to electron cooling (e.~g., interaction with the polymer matrix). On the other hand, more importantly, it suggests that sample optimization could lead to enhanced saturable absorption and, thus, much better performance and lower power consumption.

The above estimates are made under the assumption of efficient relaxation due to carrier-carrier collisions, both intraband and interband. However, in the case of linear dispersions near the Dirac point (i.e., for SLG), pair carrier-carrier collisions cannot lead to interband relaxation, thereby conserving the total number of electrons and holes separately. A three-particle collision is required to move an electron from the conduction to the valence band~\cite{Guinea96,FosterAleiner2009}. Interband relaxation by phonon emission can occur only if the electron and hole energies are close to the Dirac point (within the phonon energy). Note that for graphite and graphite flakes the situation is completely different: the dispersion near the Dirac point is quadratic, and pair carrier collisions can lead to interband relaxation. Thus, in principle, decoupled graphene monolayers (as in our present experiment, as shown by the FLG Raman spectra) will provide the highest saturable absorption for a given amount of material. To precisely estimate such effect, three-particle collisions would have to be considered. In our case, interband relaxation due to phonon emission seems insufficient to balance the pumping of electrons and holes at the rate given by Eq.~(\ref{Jin=}) (see Methods). Thus, three-particle collisions do play a role. This is not surprising, since the dimensionless Coulomb coupling constant $r_s$ is not very small, and the effective temperature, $T_{eff}=0.13\:\mbox{eV}$, is quite high for the corresponding time scale, $(r_s^4T_{eff})^{-1}$, to be in the femtosecond range.
\begin{figure}
 \centerline{\includegraphics[width=90mm]{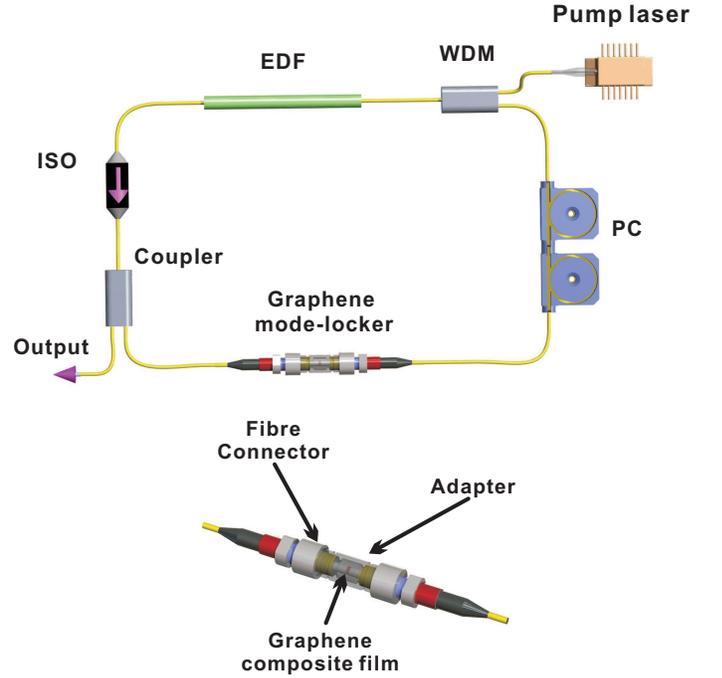}}
 \caption{\label{setup} Graphene mode-locked fibre laser and mode-locker assembly. ISO: isolator; WDM: Wavelength division multiplexer; PC: Polarization controller;EDF: Erbium doped fibre.}
\end{figure}

We now use our composite to build and test an ultrafast laser working at the main telecommunications window. The mode-locker is assembled by sandwiching the graphene-PVA between two fibre connectors with a fibre adapter, as schematized in Fig.\ref{setup}. A 0.8m Erbium doped fibre (EDF) is used as the gain medium. It is pumped by a 980nm diode laser via a wavelength-division-multiplexer (WDM). An isolator (ISO) is placed after the gain fibre to maintain unidirectional operation. A polarization controller (PC) optimizes mode-locking. A 20/80 coupler is used. The total cavity length is 10.54~m. The input mode diameter is $\sim$10~$\mu$m.
\begin{figure*}
 \centerline{\includegraphics[width=180mm]{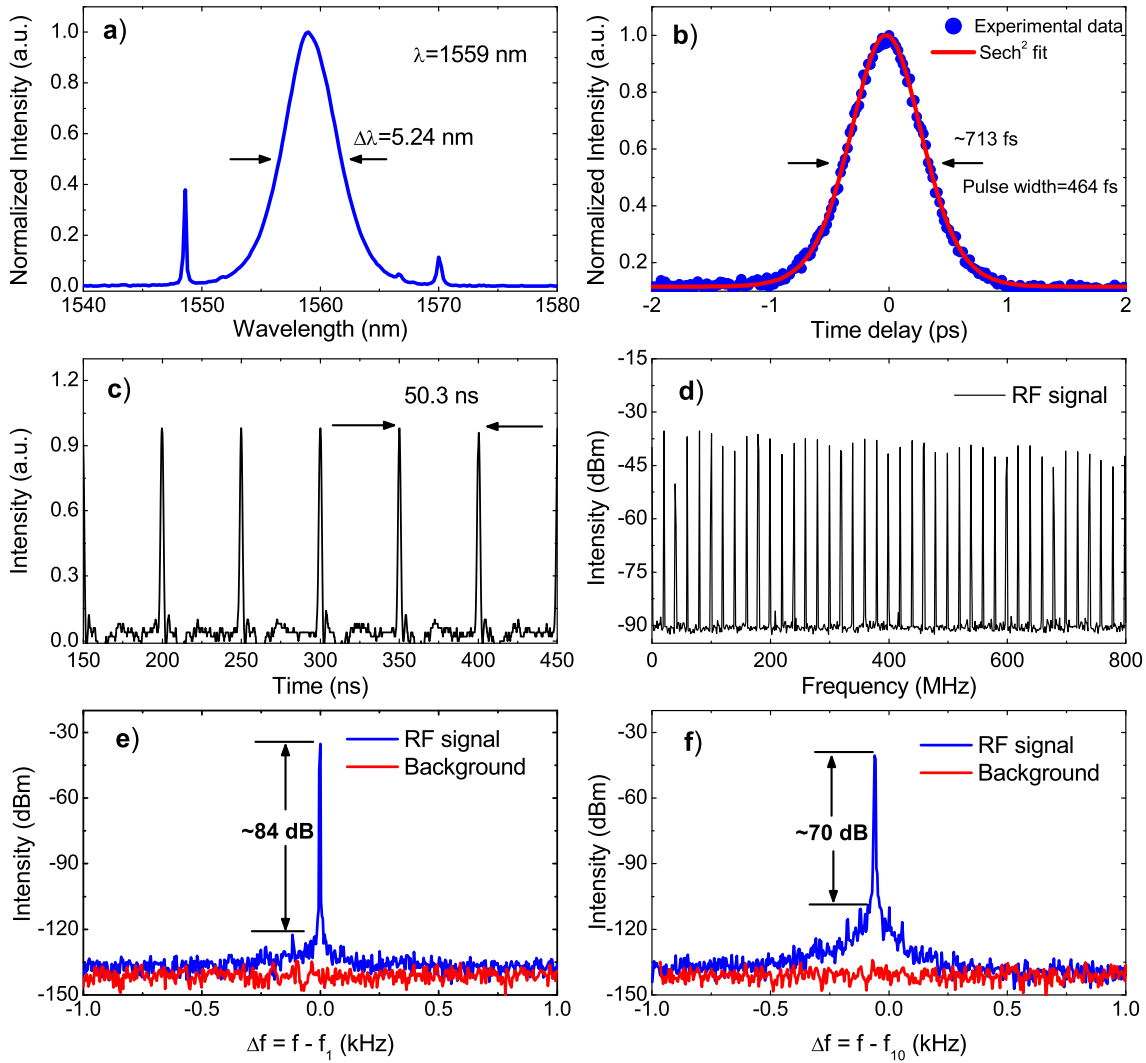}}
  \caption{\label{Laser_res}Mode-locked pulses characteristics.(a)Output spectrum; Spectral resolution 0.1nm.(b)Autocorrelation trace of output pulses;Delay resolution$\sim$20fs.(c)Oscilloscope trace.(d)Wideband RF spectrum up to 0.8GHz.(e)RF spectrum measured around the fundamental repetition rate, $f_{1}$=19.9 MHz.(f)RF spectrum measured around the tenth harmonic of the repetition rate, $f_{10}$=199MHz. In (e,f), the measurement span is 2KHz with 1-Hz resolution bandwidth. The red trace depicts the background when the laser is switched off.}
\end{figure*}

The threshold pump power for continuous wave lasing is $\sim$10mW. When this is increased to $\sim$27mW, stable mode-locking can be initiated by introducing a disturbance to the intracavity fibre. Once stable output is achieved, no further polarization controller adjustment is needed. It is always possible to decrease the pump power to $\sim$22mW while maintaining mode-locking. When mode-locked, the laser produces a pulse train at a rate of 19.9MHz.

Fig.\ref{Laser_res}(a) shows a typical output spectrum, with central wavelength $\sim$1559nm. The full width at half maximum (FWHM) bandwidth is 5.24nm. The sidebands at 1548.6 and 1570nm, are typical of soliton-like pulse formation~\cite{Dennis1994}, resulting from intracavity periodical perturbations~\cite{Dennis1994}. Fig.\ref{Laser_res}(b) is a second harmonic generation (SHG) autocorrelation trace of the mode-locked pulses after a 50cm single mode fibre lead from the coupler, with FWHM=713 fs. Assuming a $sech^{2}$ temporal profile, the deconvolution gives a pulse duration of 464fs.  The time-bandwidth product (TBP) is 0.3. Due to the autocorrelator delay resolution (20fs,5\%), there is a minor deviation from 0.315, expected for transform-limited $sech^{2}$ pulses, having the shortest duration for a given spectral width. This confirms soliton-like operation\cite{AgrawalBook}. Figure~\ref{Laser_res}(c) is the output pulse train, with a period 50.3ns, as expected from the fibre cavity length. To study the operation stability, we measure the radio-frequency (RF) spectrum. We first perform wide span frequency spectrum measurements up to 800MHz, as in Fig.\ref{Laser_res}(d). This shows no significant spectral modulation, implying no Q-switching instabilities\cite{Honninger1999}. Figs.\ref{Laser_res}(e,f) are the RF spectrum around the fundamental repetition rate and its tenth harmonic. A $>$70dB peak-to-background ratio ($10^7$ contrast) is observed, indicating good mode-locking stability\cite{Von1986}.

Our graphene-based ultrafast laser, harnessing the wideband optical nonlinearity of graphene, with no need of bandgap engineering, extends the practical potential of this novel material from nanoelectronics to optoelectronics and integrated photonics.

We acknowledge funding from EPSRC GR/S97613/01, EP/E500935/1, the European Research Council grant NANOPOTS, a Royal Society Brian Mercer Award for Innovation, the Isaac Newton Trust and the Cambridge Integrated Knowledge Centre in Advanced Manufacturing Technology for Photonics and Electronics.

\section*{Methods}
\subsection{Sample Preparation and Characterization}
Dispersions are prepared by ultrasonicating (Branson 450A, 180W power, 8-10$^{o}$C) 120mg graphite (Sigma Aldrich) for 2hrs in 10ml deionized (DI) water, with 90mg Sodium deoxycholate (SDC) bile salt. The resulting dispersion is then centrifuged at 5 krpm (2500\textit{g}) using a TH-641 swing rotor in a Sorvall WX-100 Ultra centrifuge for 1hr at 18$^{o}$C. The PVA-graphene composite is prepared by mixing 4ml of the centrifuged dispersion with 2ml 15wt\% aqueous PVA (Wako chemicals) solution in a Hauschild Speedmixer (DAC150 FVZ-K) for 5min at 3krpm. The mixture is then vacuum evaporated at 20$^{o}$C, resulting in a $\sim$50$\mu$m thick composite.

For UV-Vis-NIR absorption spectroscopy, 1 part of centrifuged dispersion with 10 parts of 0.9wt\% SDC-water solution are mixed. The absorption is background subtracted to account for solvent and surfactant. A Perkin-Elmer Lambda 950 UV-vis spectrometer is used. The centrifuged dispersions are diluted as before and drop-cast onto a Si wafer with 300nm thermally grown SiO$_{2}$. The drops are dried at 20$^{o}$C. The samples are then rinsed in acetone for 5mins and water for an additional 5 mins. This removes the surfactant. These samples are then used for Raman and SEM characterization. Raman spectra are collected with a Renishaw 1000 spectrometer at 514.5nm using a 100x objective, and $\sim$1mW power.

For power-dependent transmission measurements, the graphene mode-locker assembly is coupled to a $\sim$580fs source, tuneable from 1548 to 1568 nm. The typical spectral bandwidth at 1558nm is 2.8nm. This is achieved by filtering a homemade SWNT fibre laser ($\sim$400 fs, 38.83MHz) using a 3-nm band-pass filter. The laser beam is then amplified by an Erbium-doped fibre amplifier and a 20\% tap monitors the input power. Another 80\% is used to pump the graphene mode-locker assembly. Two calibrated power-heads are programmed to read the input/output power simultaneously. The estimated effective mode area on the composite, $A_{eff}$, is $78.54\:\mu\mbox{m}^2$, deduced from the $\sim10\mu$m mode diameter.

A spectrum analyzer (Anritsu MS9710B) and a SHG autocorrelator (APE Pulsecheck 50) are used to measure the output spectrum and pulse width. The RF spectrum is measured using a home-made lightwave converter connected to a spectrum analyzer (Anritsu MS2719B).

For a pump wavelength of 1558~nm, the total number of pump photons per cm$^{2}$ per pulse ($\Gamma_{tot}$) at 5.35~mW is
\[
\Gamma_{tot}=\frac{5.35\:\mbox{mW}}{38.83\:\mbox{MHz} \times  E_L \times A_{eff}} = 1.38 \times 10^{15}\:\mbox{cm}^{-2}.
\]
As discussed in the main text, the measured transmission of a packaged reference pure PVA device is $\tau_0=82.6\%$. Considering the absorption of $N$ graphene layers, $1-(1-\alpha_1)^N$, with $\alpha_1=2.3\%$, and linear coupling losses, $1-\tau_0$, to cause the non-saturated loss $1-\tau$ ($\tau=65.7\%$), the total number of photons absorbed by the SLG and FLG flakes per cm$^{2}$ per pulse for our maximum pump in the power-dependent absorption measurements is $(82.6 \%-65.69 \%)\times\Gamma_{tot} = 2.3 \times 10^{14}\:\mbox{cm}^{-2}$. This can be estimated to be on average equivalent to the effect of $\sim$11 graphene layers: $65.7/82.6\sim(1-2.3\%)^{10.7}$. Then, the average number of photons absorbed per layer per cm$^{2}$ per pulse, is $\sim(2.3\times10^{14}\:\mbox{cm}^{-2})/{11}= 2.2 \times 10^{13}\:\mbox{cm}^{-2}$ .

\begin{widetext}

\subsection{Kinetic Equations}
The kinetic equations for the evolution of electron and hole distribution functions $f_e(\vec{p}),f_h(\vec{p})$ due to phonon emission can be written analogously to Ref. 27. Here the task is even simpler as (i)~any spatial dependence is absent, (ii)~due to frequent collisions each photoexcited electron-hole pair is assumed to quickly forget its initial state, so the joint electron-hole distribution function $f(\vec{p},\vec{p}')$ factorizes into a product $f_e(\vec{p})\,f_h(\vec{p}')$. For $f_{e,h}(\vec{p})$ we have:
\begin{eqnarray}
&&\frac{\partial{f}_e(\vec{p})}{\partial{t}}=
\int\frac{v^2d^2\vec{p}'}{(2\pi)^2}\left\{R(\vec{p}',\vec{p})\,f_e(\vec{p}')[1-f_e(\vec{p})]
-R(\vec{p},\vec{p}')\,f_e(\vec{p})[1-f_e(\vec{p}')]\right\}
-\int\frac{v^2d^2\vec{p}'}{(2\pi)^2}\,
\tilde{R}(\vec{p},\vec{p}')\,f_e(\vec{p})\,f_h(\vec{p}'),\nonumber\\\\
&&\frac{\partial{f}_h(\vec{p})}{\partial{t}}=
\int\frac{v^2d^2\vec{p}'}{(2\pi)^2}\left\{R(\vec{p}',\vec{p})\,f_h(\vec{p}')[1-f_h(\vec{p})]
-R(\vec{p},\vec{p}')\,f_h(\vec{p})[1-f_h(\vec{p}')]\right\}
-\int\frac{v^2d^2\vec{p}'}{(2\pi)^2}\,
\tilde{R}(\vec{p},\vec{p}')\,f_e(\vec{p})\,f_h(\vec{p}').\nonumber\\
\end{eqnarray}
The phonons are assumed to be at zero temperature. The interband relaxation kernel $R(\vec{p},\vec{p}')$ is a sum of three terms, corresponding to emission of (i)~optical phonons with wave vectors near $K$, (ii)~optical phonons with wave vectors near $\Gamma$, and (iii)~longitudinal acoustic phonons:
\begin{eqnarray}
&&R(\vec{p},\vec{p}')=R_K(\vec{p},\vec{p}')
+R_\Gamma(\vec{p},\vec{p}')+R_\mathrm{ac}(\vec{p},\vec{p}'),\\
&&R_K(\vec{p},\vec{p}')=
2\lambda_K\sin^2\frac{\varphi_{\vec{p}}-\varphi_{\vec{p}'}}{2}\,
\pi\delta(vp-vp'-\omega_K),\\
&&R_\Gamma(\vec{p},\vec{p}')=\lambda_\Gamma\,\pi\delta(vp-vp'-\omega_\Gamma),\\
&&R_\mathrm{ac}(\vec{p},\vec{p}')=\ell_\mathrm{ac}|\vec{p}-\vec{p}'|\,
\cos^2\frac{\varphi_{\vec{p}}-\varphi_{\vec{p}'}}{2}\,
\pi\delta(vp-vp'-v_s|\vec{p}-\vec{p}'|).
\end{eqnarray}
The interband kernel contains only the optical phonon contribution:
\begin{eqnarray}
&&\tilde{R}(\vec{p},\vec{p}')=\tilde{R}_K(\vec{p},\vec{p}')
+\tilde{R}_\Gamma(\vec{p},\vec{p}'),\\
&&\tilde{R}_K(\vec{p},\vec{p}')=
2\lambda_K\cos^2\frac{\varphi_{\vec{p}}-\varphi_{\vec{p}'}}{2}\,
\pi\delta(vp+vp'-\omega_K),\\
&&\tilde{R}_\Gamma(\vec{p},\vec{p}')
=\lambda_\Gamma\,\pi\delta(vp+vp'-\omega_\Gamma).
\end{eqnarray}
Here we use the Einstein model for the optical phonon dispersion (momentum-independent frequencies $\omega_K,\omega_\Gamma$) and the Debye model for the acoustic phonon dispersion (the frequency of the phonon with wave vector~$\vec{q}$ is $v_s|\vec{q}|$, where $v_s\ll{v}$ is the sound velocity). We introduce the dimensionless electron-phonon coupling constants $\lambda_K,\lambda_\Gamma$ for the optical phonons, while the analogous coupling constant for the acoustic phonons has the dimensionality of length, since the coupling is proportional to the phonon wave vector:
\begin{eqnarray}
&&\lambda_\Gamma=\frac{F_\Gamma^2}{Mv^2\omega_\Gamma}\frac{\sqrt{27}a^2}4,\quad
\lambda_K=\frac{F_K^2}{Mv^2\omega_K}\frac{\sqrt{27}a^2}4,\quad
\ell_\mathrm{ac}=\frac{D_0^2}{Mv^2v_s}\frac{\sqrt{27}a^2}4.
\end{eqnarray}
Here $F_K$~and $F_\Gamma$ are the force constants measuring the change in the electron energy with lattice displacement along the corresponding normal mode (in the nearest-neighbor tight-binding model $F_\Gamma=F_K=3(\partial{t}_0/\partial{a})$, where $\partial{t}_0/\partial{a}$ is the derivative of the nearest-neighbor electronic coupling matrix element~$t_0$ with respect to the bond length~$a$), $D_0$ is the deformation potential, $M$~is the mass of the carbon atom, $\sqrt{27}a^2/4$ is the area per carbon atom (half of the unit cell area). These definitions coincide with those of Ref.~27.

For a rigorous treatment of the problem, this kinetic equation should be supplemented with a source term describing the optical pumping of electron and hole populations, as well as electron-electron collision terms for both pair and triple collisions, and then solved. For an estimate, we assume the electron-electron relaxation to be efficient enough to quickly establish the Fermi-Dirac distribution~(\ref{FermiDirac=}) and calculate the cooling rate due to the phonon emission. Combining the definition of $\mathcal{E}$ [Eq.~(2)] with the kinetic equation, we get:
\begin{eqnarray}
\left.-\frac{d\mathcal{E}}{dt}\right|_{ph}&=&
4\int\frac{d^2\vec{p}}{(2\pi)^2}\int\frac{v^2d^2\vec{p}'}{(2\pi)^2}
\left\{f_e(\vec{p})\,[1-f_e(\vec{p}')
+f_h(\vec{p})\,[1-f_h(\vec{p}')]\right\}\times\nonumber\\
&&\qquad\times\left[\omega_K{R}_K(\vec{p},\vec{p}')+
\omega_\Gamma{R}_\Gamma(\vec{p},\vec{p}')
+v_s|\vec{p}-\vec{p}'|\,R_\mathrm{ac}(\vec{p},\vec{p}')\right]+\nonumber\\
&&{}+4\int\frac{d^2\vec{p}}{(2\pi)^2}
\int\frac{v^2d^2\vec{p}'}{(2\pi)^2}\,f_e(\vec{p})\,f_h(\vec{p}')
\left[\omega_K\tilde{R}_K(\vec{p},\vec{p}')+
\omega_\Gamma\tilde{R}_\Gamma(\vec{p},\vec{p}')\right].
\end{eqnarray}
Substituting here the Fermi-Dirac distribution~(\ref{FermiDirac=}) and calculating the integrals in the approximation $v_s\ll{v}$, we arrive at Eq.~(5), with $L_\mathrm{ac}=\ell_\mathrm{ac}v_s/v$.

The rate of interband relaxation due to phonon emission is calculated analogously,
\begin{equation}
\left.-\frac{dn_{e,h}}{dt}\right|_{ph}=4\int\frac{d^2\vec{p}}{(2\pi)^2}
\int\frac{v^2d^2\vec{p}'}{(2\pi)^2}\,f_e(\vec{p})\,f_h(\vec{p}')
\left[\tilde{R}_K(\vec{p},\vec{p}')+
\tilde{R}_\Gamma(\vec{p},\vec{p}')\right]\leq\frac{\lambda_\Gamma\omega_\Gamma^3+\lambda_K\omega_K^3}{6\pi{v}^2}.
\end{equation}
The upper limit on the relaxation rate is obtained by setting $f_e=f_h=1$. We get $9\times{10}^{12}\:\mbox{cm}^{-2}\mbox{ps}^{-1}$, which is four times smaller than the pumping rate in Eq.~(3), thus validating our assumptions.
\end{widetext}

\end{document}